\documentclass{webofc}
\usepackage[varg]{txfonts}   
\begin{document}
\title{Perspectives of Semi-Inclusive Deep-Inelastic Scattering}
%
%

\author{\firstname{Anselm} \lastname{Vossen}\inst{1,2}\fnsep\thanks{\email{anselm.vossen@duke.edu}}}

\institute{Duke University
\and Jefferson Lab}

\abstract{%
This contribution highlights some topics addressed by current and future experiments in Semi-Inclusive Deep-Inelastic Scattering. 
We concentrate on the programs at 12 and 22 GeV at Jefferson Lab using the CLAS detector, and at the future Electron-Ion Collider.   
}
\maketitle

\section{Introduction}
Semi-Inclusive Deep-Inelastic Scattering (SIDIS) is aprocess in which a leptonic probe, usually an electron, is scattered off a proton in the process $e+p\rightarrow h_1 + X$. Part of the final state is detected, here hadron $h_1$. 
The SIDIS process has been instrumental to reveal the quark-gluon substructure of the proton and its dynamics. At leading twist, the SIDIS cross-section can be described by the combination of parton distribution functions (PDFs) and fragmentation Functions (FFs) as well as the elementary hard scattering cross-section of the partons. PDFs and FFs encode the non-perturbative aspects of initial and final states, respectively. At leading twist, they can be interpreted in a probabilistic way. Details can be found in the recent reviews~\cite{Metz:2016swz, Anselmino:2020vlp}. We follow the notatation also used in Ref.~\cite{Anselmino:2020vlp}, i.e. the collinear unpolarized, helicity and transversity parton distribution functions are called $f$, $g$ and $h$, respectively. Analogously, the fragmentaiton function are called $D$, $G$ and $H$ with a superscript $\perp$ indicating transverse momentum dependence, a subscript $1$ indicates leading twist.
Naively, the different PDFs can be extracted from azimuthally dependent SIDIS cross-section as they appear in the expression at leading twist as combinations with fragmentation functions as well as kinematic factors that can be approximatly described by functions of $y$, the fractional energy transfer from the scattered electron.
For example, the unpolarized pdf appears with the unpolarized FF as a term 
$(1+(1-y)^2) \sum_{q,\bar{q}} e_q^2f_1^q(x)D_1^q(z,P_{k_\perp})$, whereas the Sivers TMD appears as $|S_T|(1-y)P_{h\perp}/zM_h  \sin(\phi_h-\phi_S)\sum_{q,\bar{q}}e_q^2 f_{1T}^{\perp,q}(x)D_1^q(z,P^2_{h\perp})$. Here $x, z, M_h, P_{h\perp}$  are defined as usual as the bjorken scaling variable, the fractional momentum of the initial quark carried by the final state hadron, the hadron mass (in multi hadron states, this can be the invariant mass), and the transverse hadron momentum respectively.

However, the parton picture is an approximation and its validity depends on the kinematics. At lower momentum transfers $Q^2$ higher twist (e.g. twist-3) effects can contribute significantly~\cite{Anselmino:2020vlp} and the kinematics are far enough from the Bjorken limit that significant contributions from the target fragmentation or the soft region~\cite{Boglione:2019nwk} can complicate the interpretation of observables in the parton picture.
Furthermore, progress in MC modeling allows to understand hard scattering data in a model dependent but more intuitiv way, opening a new avenue for exploration~\cite{Accardi:2022oog}. MC models often succeed in parts of the phase space where the parton model assumption is not valid but tend to have many model parameters that need to be tuned to data. On the other hand, it is these parameters that can often be interpreted in terms of a descriptive physics model.
To disentangle these contribution to the cross-section it is important to collect data over a wide range of kinematics (e.g. $Q^2$-$x$ reach) with high luminosity. Additionally, targets with different polarization and different quark content (e.g. $p$ and $He^3$) are needed to access the different polarized effects and do a flavor separation.
This contribution highlights some future measurements that will be of interest at current and future facilities.

\section{Experimental Perspectives}
With the upgrade to 12~GeV, CEBAF at Jefferson Lab~\cite{Accardi:2023chb} can provide luminosities up to  $10^{38}$cm$^{-2}$s$^{-1}$ which is about a factor of 10$^6$ larger than the previous HERMES and COMPASS experiments. While CEBAF delivers beam to several experimental halls, here we discuss only the physics program with the CLAS detector as it is the only general purpose detector with the large acceptance needed to do precision SIDIS studies~\cite{Burkert:2020akg}. CLAS has taken data with unpolarized proton and deuterium data as well as longitudinally polarized protons. Plans to take data in the near future on transversely polarized protons and $He^3$ exist~\cite{He3Proposal}. While currently CLAS cannot accept the full luminosity at JLab, limited upgrades can help. It should also be mentioned that the proposed SoLID experiment~\cite{JeffersonLabSoLID:2022iod} is designed to take advantage of the full luminosity at JLab. 
Future proposed upgrades at JLab would allow taking data at higher energy with an upgrade of the CEBAF energy to 22~GeV~\cite{Accardi:2023chb}.
With this, the CLAS experiment is a good place to do precision measurements of the SIDIS process. However, due to the limited beam energy, a significant part of the phase space is not accessible at CLAS. Most obviously measurements at low-$x$ are not possible, due to the $x$-$Q^2$ correlation. But also at high $x$ the need to exclude final states with low invariant mass $W$ limits precision. Furthermore, the low beam energy limits the applicability of perturbative QCD~\cite{Boglione:2019nwk}.

\begin{figure}[ht]
    \centering
    \includegraphics[trim=0.5cm 10cm 0.5cm 10cm ,clip,width=0.6\textwidth]{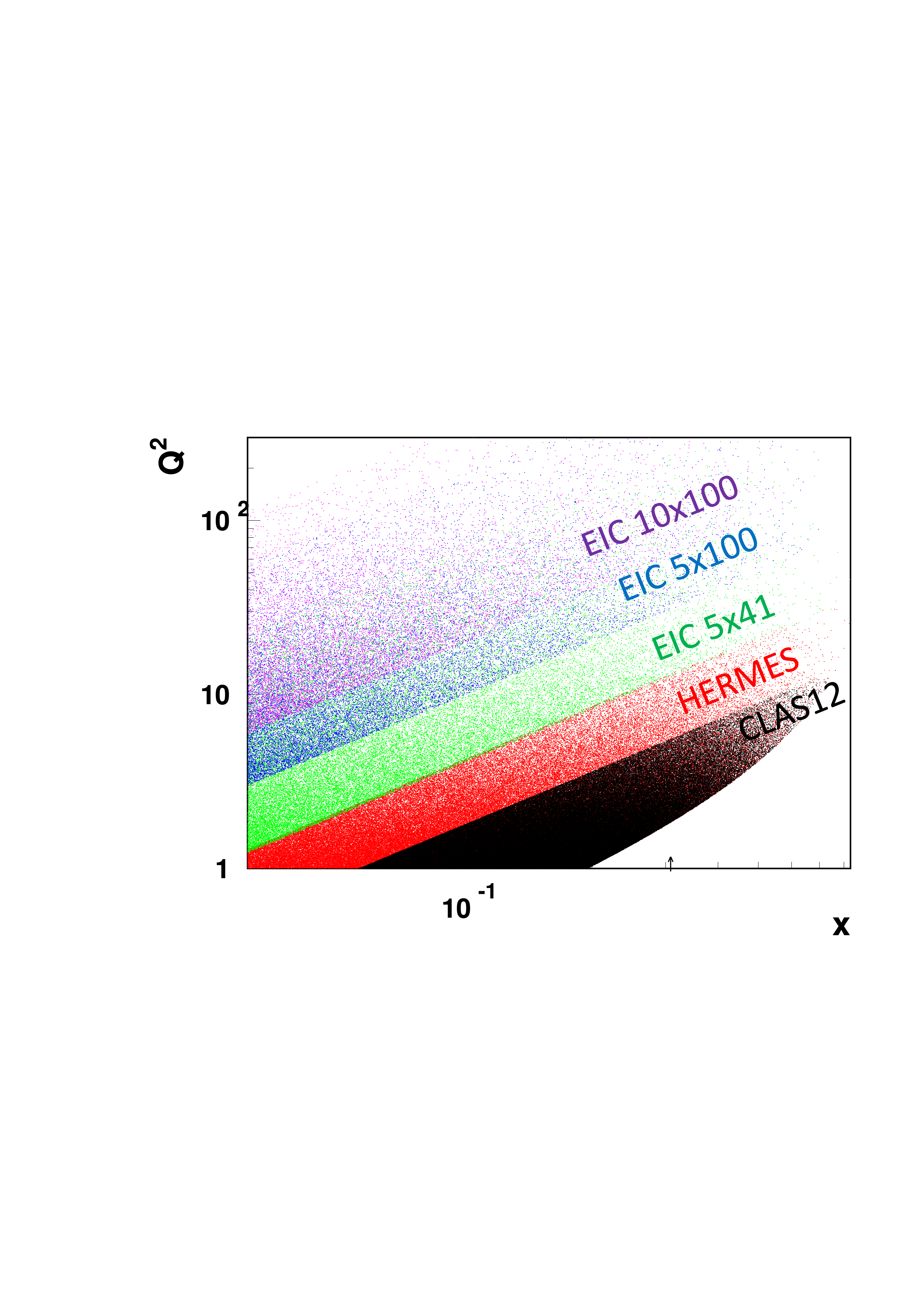}
    \caption{Estimated coverage of JLab12, HERMES and EIC data for different energy configurations. The data are constrained to $y>0.05$. Figure from Ref.~\cite{Burkert:2022hjz}}
    \label{fig:kinematics}
\end{figure}

The advent of the EIC~\cite{AbdulKhalek:2021gbh} allows to complement the JLab data with collider data at higher center-of-mass energies between 20 and 140 GeV. However, given the collider configuration the instantaneous luminosity is lower, currently projected to be between 0.5$\times$ 10$^{33}$ and 10$^{34}$ cm$^{-2}$s$^{-1}$, depending on the center of mass energy. The luminosity at the lower end is limited by beam-beam effects in the hadron beam whereas at the higher energies bremsstrahlung of the electron beam is a limitation~\cite{Willeke:2021ymc}. The kinematic regions covered by JLab and the EIC are shown in Fig.~\ref{fig:kinematics}. Here, a limit on $y$ of 0.05 is imposed, corresponding to the limitations in the reconstruction of this variable. As shown, this leads to a limitation when accessing the valence quark region at moderate $Q^2$ where transverse momenta are not dominated by perturbative radiation.
Furthermore, the kinematics leads to a suppression of some azimuthal asymmetries in beam spin asymmetries~\cite{Burkert:2022hjz}. 

\section{Recent Results and Future Plans from CLAS}
A broad SIDIS program has be implemented at CLAS, here we will just highlight some exemplary measurements. 
Single hadron beam spin asymmetries with an unpolarized target have been measured~\cite{CLAS:2021opg}. While these asymmetries show the precision that can be reached with CLAS. The single pion asymmetries are sensitive to the twist-3 PDF $e(x)$ of which the first moment can be interpreted as the transverse force exerted by the gluon fields in an unpolarized proton on a transversely polarized struck quark while it traverses the target remnant~\cite{Burkardt:2008ps}. 
However, the respective structure function $F_{LU}^{\sin\Phi_h}$ receives contribution from four terms at subleading twist. The underlying reason is that with a single, unpolarized, hadron in the final state the number of observables that can be built is limited.
Therefore it makes sense to use final states with more degrees of freedom~\cite{Vossen:2020lfd,Vossen:2018nkg} such as hadron pairs or hadrons that carry polarization degrees of freedom like $\Lambda$ hyperons.
While the phase space for $\Lambda$ production in the current region is limited at Jlab, a vibrant program for di-hadron production exists~\cite{CLAS:2022sqt,Dilks:2021nry,Dilks:2022wyj,Hayward:2021psm, dihadronLongProposal}. 
Due to the additional degrees of freedom, the term sensitive to $e(x)$ only appears with one other term in the cross-section that is assumed to be small and that can also be determined with future measurements with a longitudinal target~\cite{Courtoy:2022kca}. Due to the similarity between polarized lambda's and di-hadrons, that can carry orbital angular momentum, it is also not surprising that $e(x)$ can be accessed in beam spin asymmetries with transversely polarized $\Lambda$'s~\cite{McEneaney:2022bsf}.
However, in di-hadrons an unlimited set of orbital angular momenta can be produced that can be extracted via a partial wave decomposition~\cite{Pisano:2015wnq}. A full partial wave analysis in SIDIS and $e^+e^-$ will give insight into spin-momentum correlations in hadronization~\cite{Accardi:2022oog}. Recent work on modeling the hadronization of di-hadrons~\cite{Kerbizi:2023cde} show encouraging consistency with data and will lead to a better understanding of the microscopic processes in hadronization.

\subsection{Longitudinal Program}
\begin{figure}[ht]
    \centering
    \includegraphics[width=0.4\textwidth]{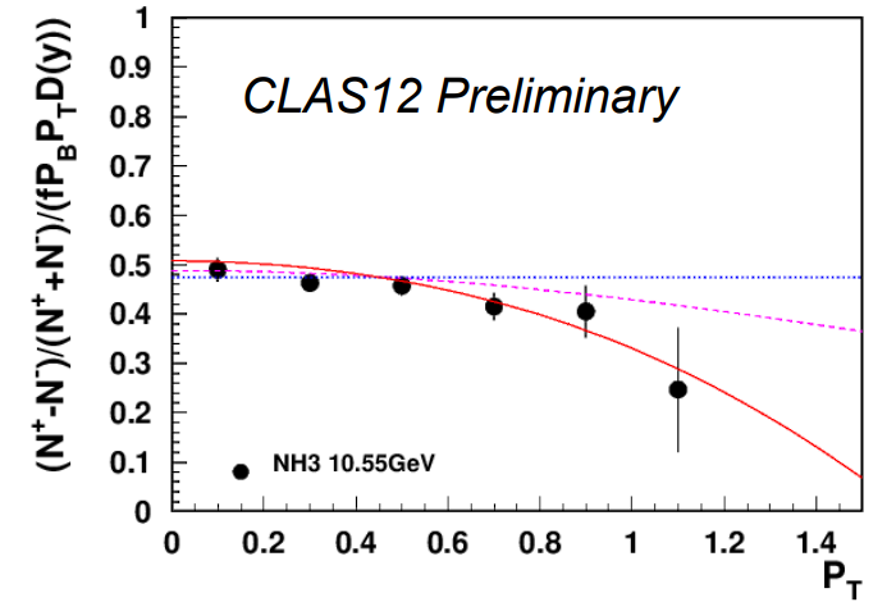}
    \caption{Preliminary longitudinal double spin asymmetries measured at CLAS12 with a polarized proton target. The asymmetries are shown vs. the transverse momentum of the hadron, thus accessing the largely unknown transverse momentum dependence of the helicity structure of the proton. Only 5\% of the available data has been used. Plot from Ref.~\cite{harutsTalk}\label{fig:A_LL}.}
    \label{fig:longTarget}
\end{figure}

Recently, a significant dataset with a longitudinal polarized target was collected at CLAS. While the analysis of this dataset is ongoing, Figure~\ref{fig:longTarget} shows a first, preliminary result on the longitudinal double spin asymmetry illustrating the potential of this dataset. Ref.~\cite{gregsTalk} contains more detail on the program.

\subsection{Fracture Functions}
Finally, with two hadrons in the final state one can select pairs combining a hadron from the target fragmentation region with a hadron in the current fragmentation function. The target fragmentation region is sensitive to fracture function while the hadron in the current fragmentation region enables sensitivity to spin correlations between target and current regions as well as to the flavor of the fragmenting quarks.
A first measurement of target and current correlation has been performed at CLAS recently~\cite{CLAS:2022sqt}. The first measurement was for beam spin asymmetries of protons and pions and a program with different targets and final state hadrons, such as $\Lambda$-$K$ is underway.

\subsection{Future Program}
\begin{figure*}
        \centering\includegraphics[width=0.25\textwidth,angle=-90]{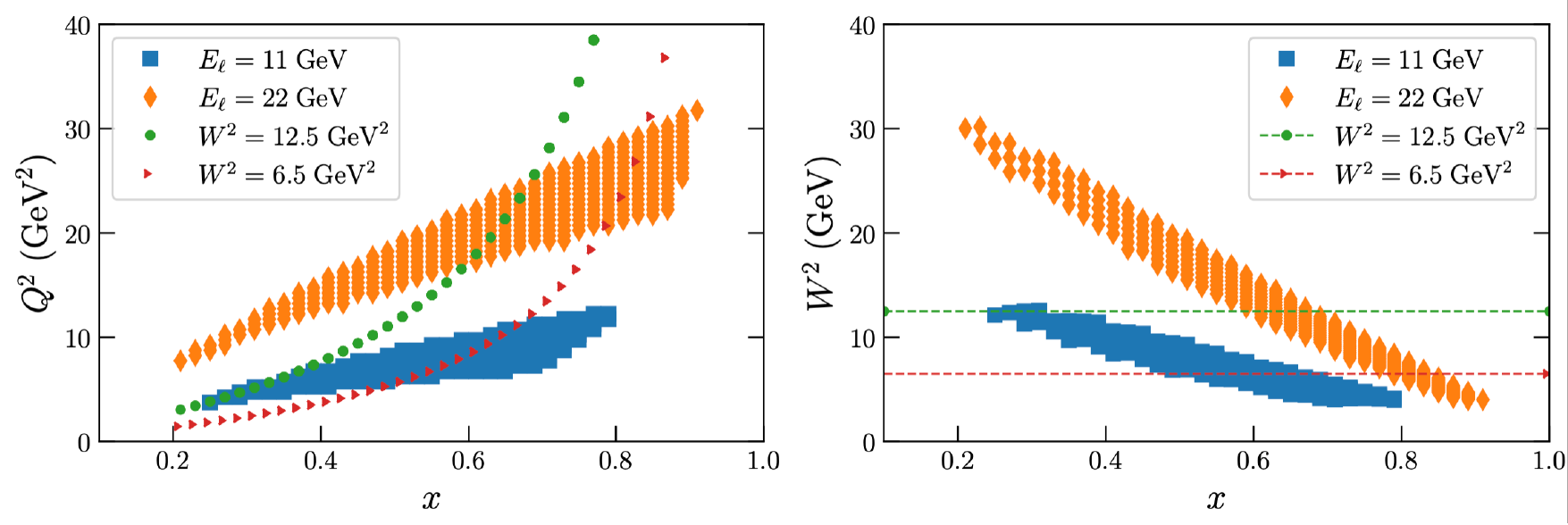}
\caption{Comparison of the kinematics covered at JLab with a 11~GeV vs a 22~GeV beam. The higher beam energy allows access to higher values of $x$ after accounting for the requirement of a minimum value of the hadronic mass $W$ to exclude the resonance region. Figure from Ref.~\cite{Accardi:2023chb}.\label{fig:clas22HighX}}
\end{figure*}

CLAS has a rich program using longitudinally and transversely polarized proton, deuterium and He3 targets as has been laid out in recent PAC proposals~\cite{transversePacProposal,He3Proposal}. Furthermore, an exciting prospect is the planned upgrade of CEBAF to 22 GeV enabling an exciting physics program outlined in a recent whitepaper~\cite{Accardi:2023chb}.
Here, we only want to highlight the prospect of accessing high $x > \approx 0.5$. This is a region relatively unexplored in SIDIS. As shown in Fig.~\ref{fig:clas22HighX}, at lower beam-energies, resonances constrain the accessible $x$ values. Previous experiments with higher beam energies were luminosity limited.
\section{A Selection of Planned Measurements at the EIC}
The physics program of the EIC has been described in detail in the recent Yellow Report~\cite{AbdulKhalek:2021gbh} and in the proposals for the Athena~\cite{ATHENA:2022hxb}, Ecce~\cite{Adkins:2022jfp} and CORE~\cite{CORE:2022rso} experiments.

As shown in Fig.~\ref{fig:kinematics}, the EIC will be instrumental in exploring the low $x$ regime. Figure~\ref{fig:A_LL} shows statistical projection for the extraction of $A_{LL}$ for kaons compared with the uncertainties of the theory which is driven by the largely unknown strange quark helicities at low $x$. As shown, the EIC will be able collect significant statistics at low $x$ with precise PID and will therefore be able to significantly constrain these distributions. 

\begin{figure*}
\centering
\includegraphics[width=0.65\textwidth]{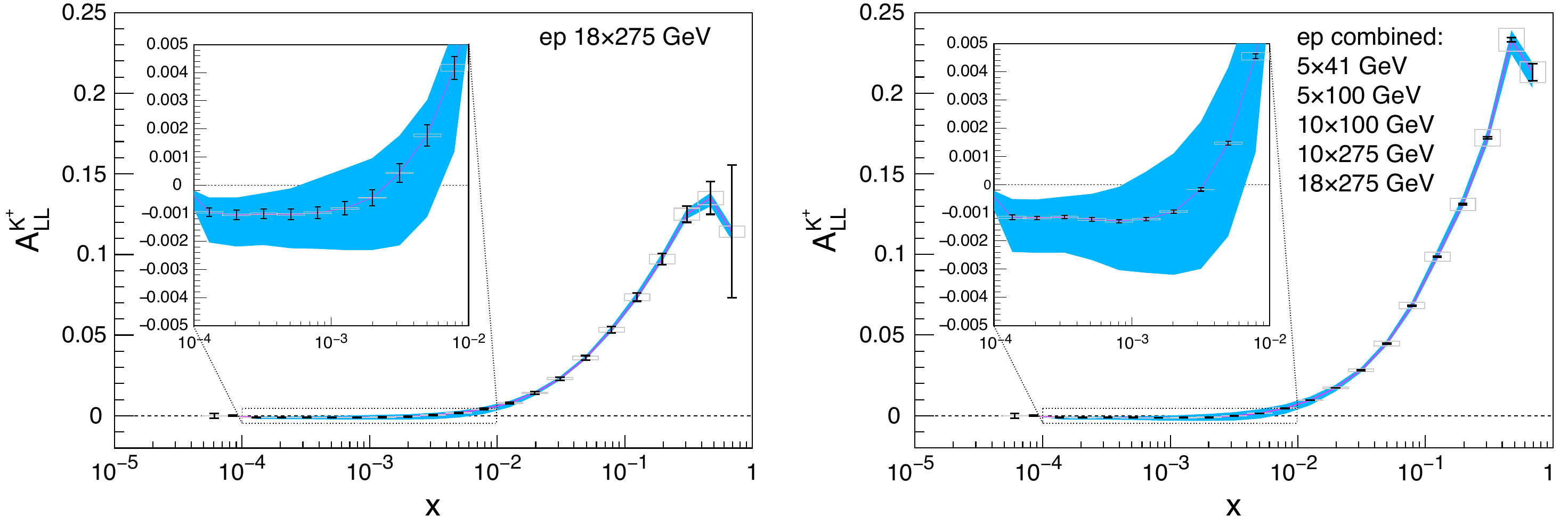}
\includegraphics[width=0.33\textwidth]{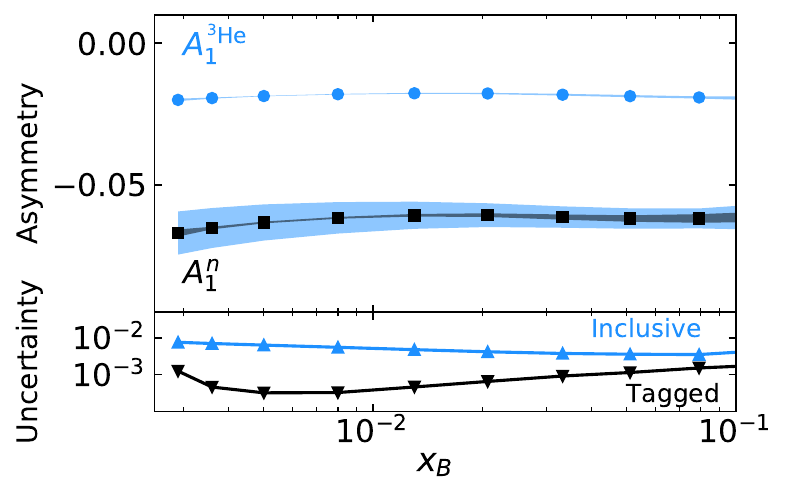}
\caption{Left/Middle: Projection for the longitudinal double spin asymmetry at the EIC using the Athena~\cite{ATHENA:2022hxb} detector for kaons using 10fb$^{-1}$ at the indicated beam energies. The comparison with the theory projections (blue band) shows the anticipated impact of the data. Figures from Ref.~\cite{ATHENA:2022hxb}. Right: Anticipated uncertainty on the extraction of the asymmetry $A_1^n$ from for neutrons using $He^3$  and spectator double tagging. The bands show the uncertainties due to nuclear corrections with and without tagging. The double tagging reduces the uncertainties by almost an order of magnitude. Figure from Ref.~\cite{Friscic:2021oti}.\label{he3tagging}}
\end{figure*}

\begin{figure*}
\centering
\includegraphics[width=0.7\textwidth]{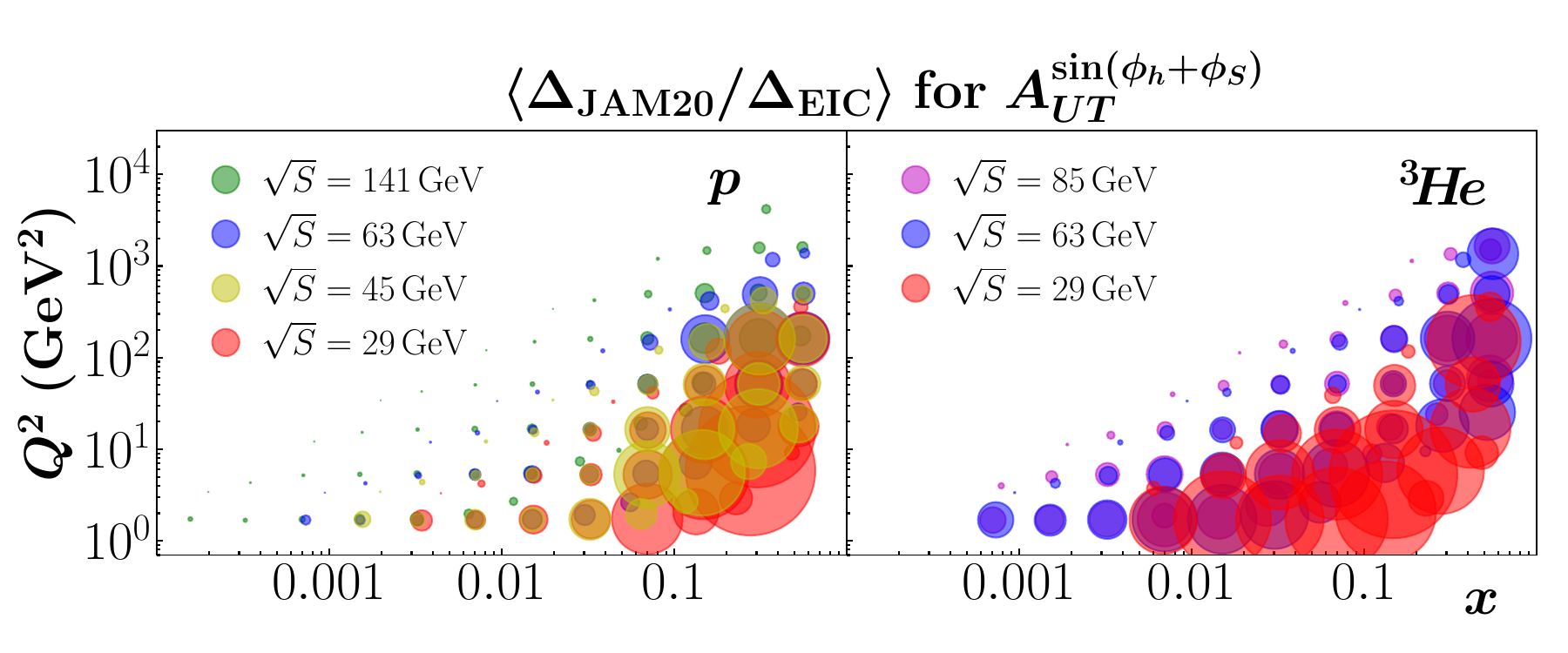}
\caption{Relative impact of the EIC data at different phase space points on the extraction of Collins asymmetries. The largest relative impact comes from the valence quark region at intermediate $Q^2$. This region is covered by lower values of $\sqrt{s}$. Figure from Ref.~\cite{Gamberg:2021lgx}.\label{fig:collinsErr}}
\end{figure*}

As mentioned above and described in detail in Ref.~\cite{Burkert:2022hjz}, a drawback of the EIC is that the kinematics suppress contributions from wormgear in longitudinal beam, transverse target spin asymmetries (LT) as well as asymmetries connected to $e(x)$ and $g_T(x)$ accessed in LU (U stands for unpolarized) and LT asymmetries. The reason is that the kinematic factor (depolarization factor) becomes small at EIC kinematics. Additionally access to small $y$, therefore large $x$, intermediate $Q^2$ is limited by the precision of the $y$ reconstruction. Typically a limit of 0.05 is set but there are developments to lower this limit~\cite{Pecar:2022vuo}.
Figure~\ref{fig:collinsErr} shows plots from Ref~\cite{Gamberg:2021lgx} demonstrating the impact of the EIC on the extraction of transversity which is exemplary for other TMD PDFs. As expected in the valence quark region at intermediate $Q^2$ data at low $\sqrt{s}$. However, as discussed before, the luminosities at these energies will be significantly lower than at higher energies.
Another feature at the EIC, enabled by the collider kinematics, is the ability to tag target fragments that stay close to the beampipe. This capability is enhanced for the proposed second detector at the EIC which can make use of a secondary focus.

A exemplary measurement is the double tagging of interactions with a $He^3$ beam. The neutron in a polarized $He^3$ is polarized to almost 90\%. Using double tagging of the protons, one can obtain a polarized neutron beam and a reconstruction of the initial neutron kinematics is possible, thus reducing nuclear corrections. A study of this process from Ref~\cite{Friscic:2021oti} is shown in Fig.~\ref{he3tagging}. A significant reduction in uncertainties is demonstrated.

\begin{figure*}
\centering
\includegraphics[width=0.8\textwidth]{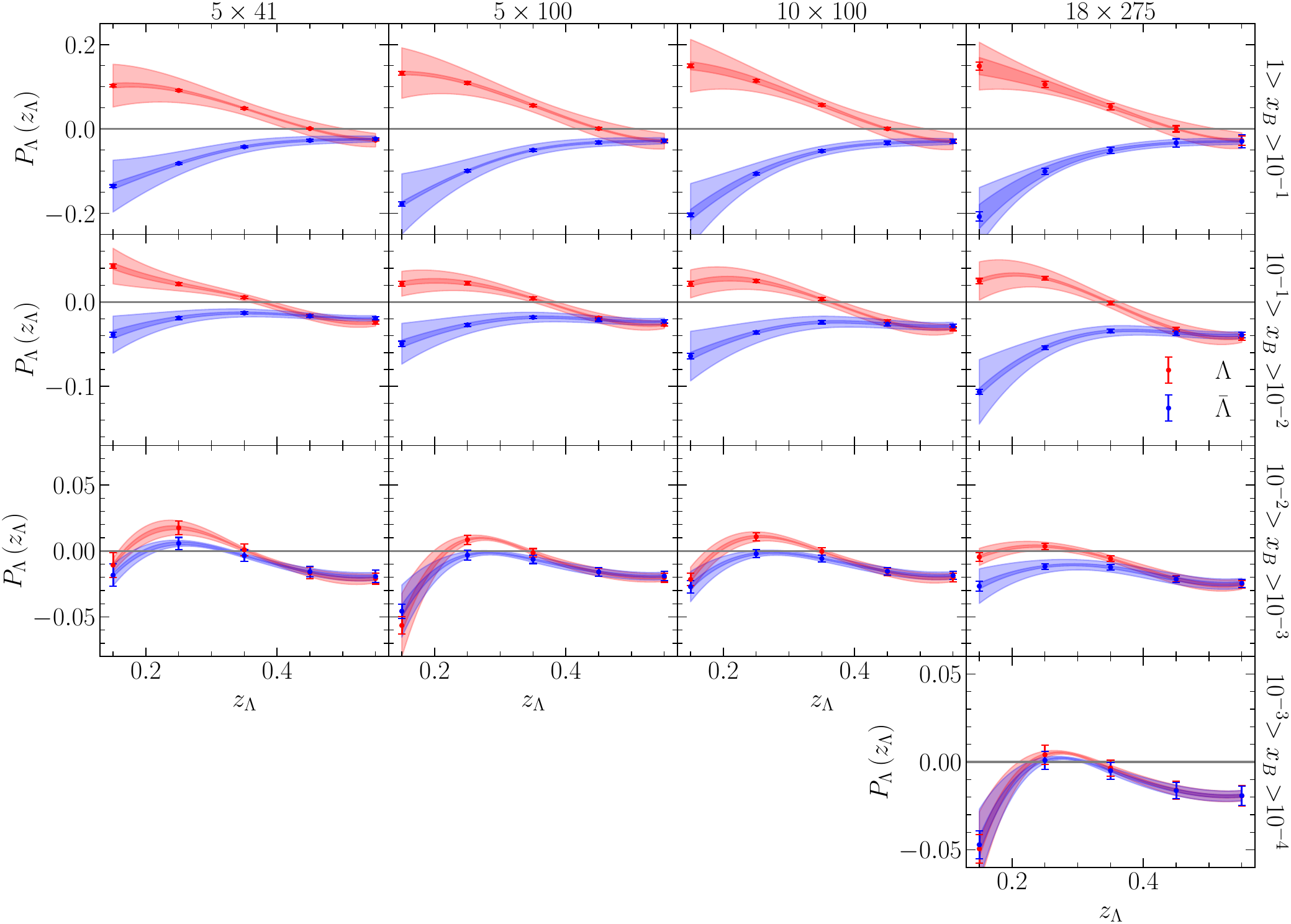}
\caption{Estimated uncertainties on the extraction of the transverse $\Lambda$ polarization compared with theory projections. The polarizing FF can be extracted from this daa. The impact of the data is evident. Figure from Ref.~\cite{Kang:2021kpt}.\label{ref:lambdaFig}}
\end{figure*}

The EIC will be the first time precision physics with polarized $\Lambda$'s in the current fragmentation region will be possible in SIDIS. Reference~\cite{Kang:2021kpt} lays out an comprehensive program for measurement of longitudinally and transversely polarized $\Lambda$'s detected semi-inclusively and in jets. 
Figure~\ref{ref:lambdaFig} showcases the the possibilities with projections for the production of transversely polarized $\Lambda$'s in unpolarized collisions as compared with theory projections. Similarly to the sign change for the Sivers asymmetry between SIDIS and Drell-Yan~\cite{Brodsky:2002cx} the comparison of the $\Lambda$ polarization between $e^+e^-$ and SIDIS is connected to the gauge structure of QCD in a fundamental way~\cite{Boer:2010ya}.

\section{Summary and Conclusion}
These contributions gave a small window on the an exciting and broad SIDIS program ongoing and planned at JLab and the future EIC.
Naturally, the space constrains for this contribution made it necessary to omit a number of exciting ongoing efforts.
One of the omissions was the COMPASS experiment, which collected additional data on the deuterium target and presented significant progrees in their analysis techniques at this meeting, e.g. in the treatment of radiative corrections.
Other topics that had to be omitted are TMDs in medium and SIDIS with charged current at the EIC.


\label{sec-1}

\bibliography{sidisPerspectives}
%
%

\end{document}